\documentclass[10pt,twocolumn,letterpaper]{article}

\usepackage{cvpr}
\usepackage{times}
\usepackage{epsfig}
\usepackage{graphicx}
\usepackage{amsmath}
\usepackage{amssymb}
\usepackage{subfig}
\usepackage{algorithm}
\usepackage{algorithmic}
\usepackage{enumitem}
\usepackage{amsthm}
\usepackage{float}
\usepackage{bbm}
\usepackage{amsmath}
\usepackage{mathtools}
\usepackage{multirow}
\usepackage{diagbox}
\usepackage{comment}
\DeclarePairedDelimiter{\norm}{\lVert}{\rVert}
\DeclarePairedDelimiter{\abs}{|}{|}
\usepackage{rotating}

\makeatletter
\def\thickhline{%
  \noalign{\ifnum0=`}\fi\hrule \@height \thickarrayrulewidth \futurelet
   \reserved@a\@xthickhline}
\def\@xthickhline{\ifx\reserved@a\thickhline
               \vskip\doublerulesep
               \vskip-\thickarrayrulewidth
             \fi
      \ifnum0=`{\fi}}
\makeatother

\newlength{\thickarrayrulewidth}
\usepackage[pagebackref=true,breaklinks=true,letterpaper=true,colorlinks,bookmarks=false]{hyperref}

 \cvprfinalcopy 

\def\cvprPaperID{7849}

\ifcvprfinal\fi
\begin{document}

\title{Bridging the Performance Gap between FGSM and PGD Adversarial Training}

\author{Tianjin Huang,Vlado Menkovski,Yulong Pei,Mykola Pechenizkiy\\

Eindhoven University of Technology\\
Eindhoven, the Netherlands\\
{\tt\small \{t.huang,v.menkovski,y.pei.1,m.pechenizkiy\}@tue.nl}

}

\maketitle

\begin{abstract}
Deep learning achieves state-of-the-art performance in many tasks but exposes to the underlying vulnerability against adversarial examples. Across existing defense techniques, adversarial training with the projected gradient decent attack (\emph{adv.PGD}) is considered as one of the most effective ways to achieve moderate adversarial robustness. However, \emph{adv.PGD} requires too much training time since the projected gradient attack (\emph{PGD}) takes multiple iterations to generate perturbations. On the other hand, adversarial training with the fast gradient sign method (\emph{adv.FGSM}) takes much less training time since the fast gradient sign method (\emph{FGSM}) takes one step to generate perturbations but fails to increase adversarial robustness. In this work, we extend \emph{adv.FGSM} to make it achieve the adversarial robustness of \emph{adv.PGD}. 
We demonstrate that the large curvature along \emph{FGSM} perturbed direction leads to a large difference in performance of adversarial robustness between \emph{adv.FGSM} and \emph{adv.PGD}, and therefore propose combining \emph{adv.FGSM} with a curvature regularization (\emph{adv.FGSMR}) in order to bridge the performance gap between \emph{adv.FGSM} and \emph{adv.PGD}. The experiments show that \emph{adv.FGSMR} has higher training efficiency than \emph{adv.PGD}. In addition, it achieves comparable performance of adversarial robustness on MNIST dataset under white-box attack, and it achieves better performance than \emph{adv.PGD} under white-box attack and effectively defends the transferable adversarial attack on CIFAR-10 dataset.    
\end{abstract}

\section{Introduction}\label{intro}
Deep Neural Networks (\emph{DNNs}) have shown great performance in  multiple tasks, e.g. image classification~\cite{Krizhevsky2012,He2016}, object detection~\cite{Girshick2014}, semantic segmentation~\cite{Long2015}, and speech recognition~\cite{Hinton2012}. However, these highly performed models show weakness on adversarial examples. Namely, carefully designed imperceptible perturbations on input can change the prediction drastically~\cite{Szegedy2013,Goodfellow2014}. This fragility prohibits \emph{DNNs} to be widespreadly applied especially in security-sensitive tasks such as autonomous cars, face recognition, and malware detection. Therefore, training a model resistant to adversarial attacks becomes increasingly important.

\begin{table}
\centering  
\begin{tabular}{lcc} \\
\thickhline
Method  &\emph{adv.PGD} &\emph{adv.FGSM}\\
\hline
\emph{PGD-l2}  &0.710 &0.353\\
\emph{PGD-inf}  &0.444 &0.091\\
\emph{Deepfool-l2($\rho_{adv}$)}  &0.178 &0.022\\
\emph{C\&W($\rho_{adv}$)}  &0.129 &0.016\\
\thickhline
\end{tabular}
\caption{Comparison of robustness performance of robust models trained by \emph{adv.FGSM} and \emph{adv.PGD} respectively against various attacks. Experiments are based on CIFAR-10 test set and \emph{ResNet-18} model. For \emph{Deepfool-l2} and \emph{C\&W-l2} attacks, $\rho_{adv}$ is calculated using Eq.~\ref{eq8}.}
\label{tb:motivation}
\end{table}

By now, plenty of ways have been proposed to generate adversarial examples, which can be categorized into black-box attack and white-box attack. White-box attack can access the complete knowledge of the target model including its parameters, architecture, training method and training data. The popular white-box attacks include \emph{FGSM}~\cite{Goodfellow2014}, \emph{PGD}~\cite{Madry2017}, \emph{Deepfool}~\cite{Moosavi-Dezfooli2016}, \emph{C\&W}~\cite{carlini2017towards}, etc. Black-box attack generates adversarial examples without knowledge of the target model, e.g. \emph{ZOO}~\cite{chen2017zoo}, \emph{Transferable adversarial attack}~\cite{liu2016delving,papernot2017practical}, etc. Correspondingly, many methods have been proposed to improve model's adversarial robustness against these attacks. Qiu et al.~\cite{Qiu2019}, Akhtar and Mian~\cite{Akhtar2018} separate these defense methods into three categories: (1) augmenting training data, e.g. adversarial training~\cite{Madry2017,Goodfellow2014}; (2) using extra tool to help model against adversarial attacks, e.g. PixelDefend~\cite{Song2017}; and (3) modifying model to improve its robustness, e.g. Defensive Distillation~\cite{Papernot2016}, Regularization~\cite{Moosavi-Dezfooli2018,jakubovitz2018improving}.

\begin{figure*}[htb]
    \subfloat[]{
        \includegraphics[width=0.24\textwidth]{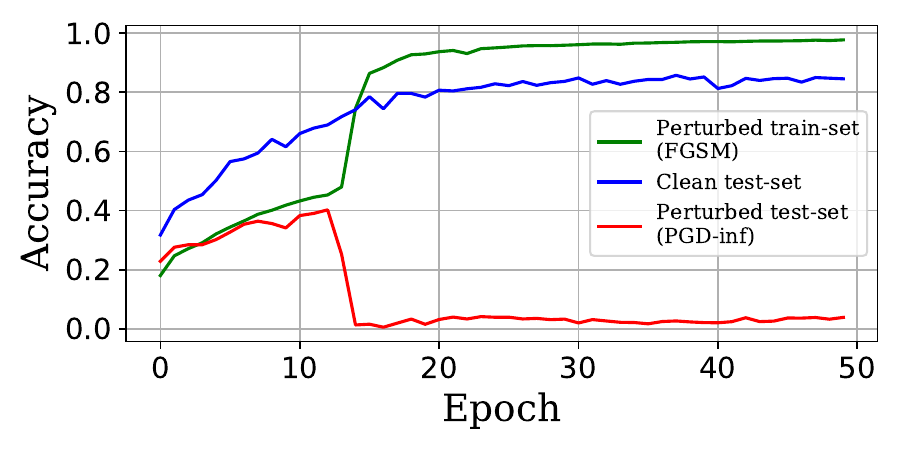}
        \label{fig:1a}
    }
    \subfloat[]{
        \includegraphics[width=0.24\textwidth]{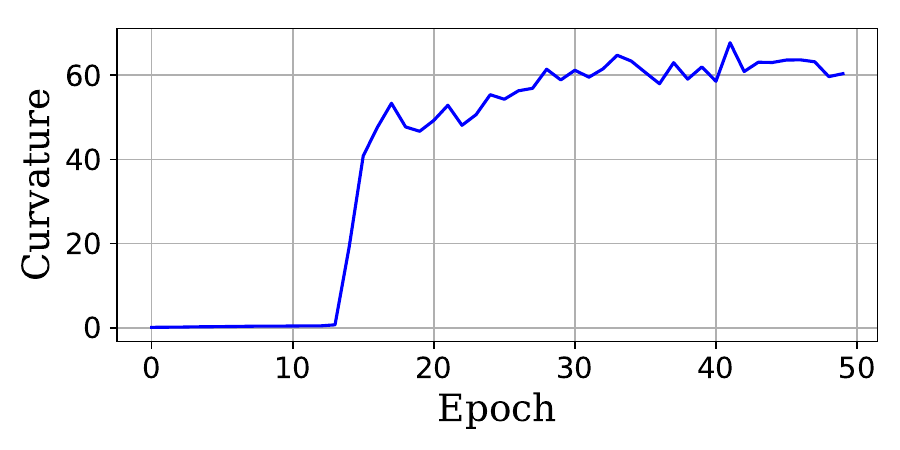}
        \label{fig:1b}
    }
    \subfloat[]{
        \includegraphics[width=0.24\textwidth]{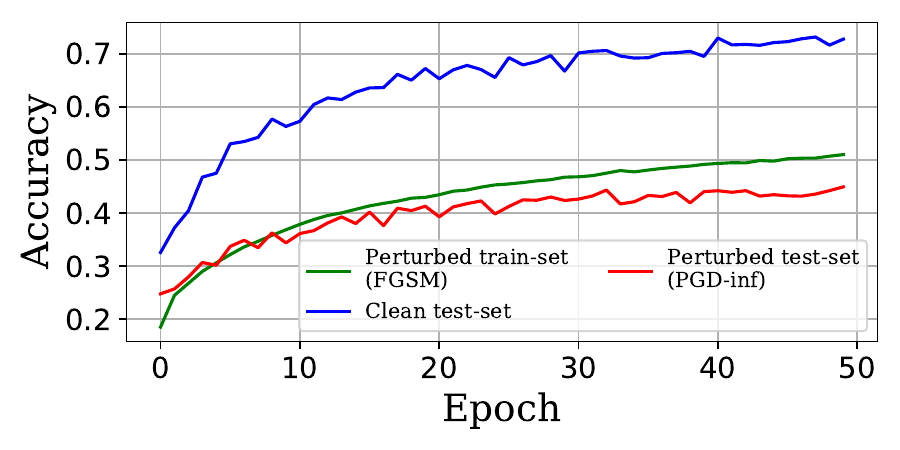}
        \label{fig:1c}
    }
    \subfloat[]{
        \includegraphics[width=0.24\textwidth]{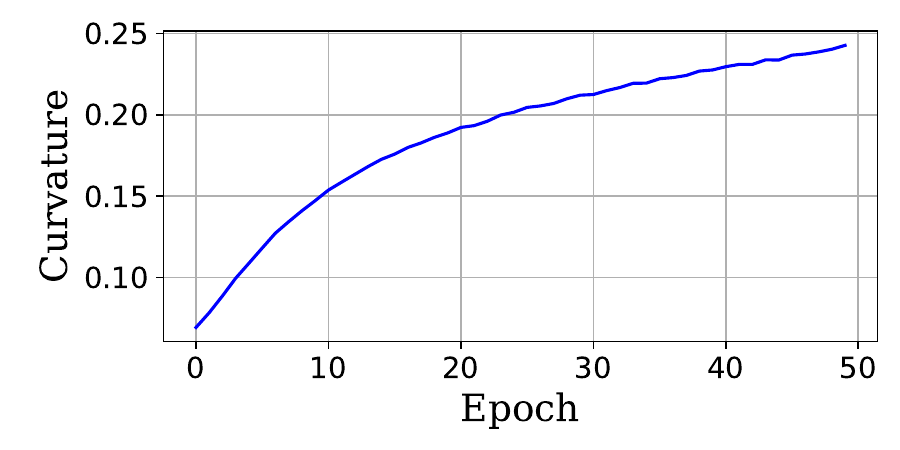}
        \label{fig:1d}
    }
\caption{The accuracy and average curvature curve for training \emph{ResNet-18} model on CIFAR-10 within 50 epochs. The subfigure \emph{(a)} and \emph{(b)} show the accuracy and average curvature curve of the model trained by \emph{adv.FGSM} respectively; \emph{(c)} and \emph{(d)} show the accuracy and average curvature curve of the model trained by our proposed \emph{adv.FGSMR} respectively. The curvature value is calculated using Eq.~\ref{eq6}. Notice: A sudden decrease of perturbed accuracy under \emph{PGD-inf} attack occurs with the sudden increase of the curvature value for \emph{adv.FGSM}.}
\label{fig:1}
\end{figure*}

Among these defense approaches, most have been reported failure on later proposed adversarial attacks except for adversarial training~\cite{athalye2018obfuscated}. \emph{adv.PGD} has been considered as one of the most effective ways to achieve moderate adversarial robustness~\cite{wang2019convergence}. However, a major issue for \emph{adv.PGD} is its expensive computational cost because \emph{PGD} attack takes multi-step iterations to generate perturbations. The high computational cost makes this method hard to be applied on larger neural networks and datasets. On the other hand, \emph{adv.FGSM} takes much less computational cost but shows no robustness improvement against adversarial attacks except for \emph{FGSM} attack (Table~\ref{tb:motivation}). The behavior of strong defense on \emph{FGSM} attack but weak defense on other attacks has also been reported in ~\cite{Madry2017,kurakin2016adversarial}. We believe that it would be of great values if we can bridge robustness performance gap between \emph{adv.FGSM} and \emph{adv.PGD}. We further explore the reasons for the lack of adversarial robustness performance of \emph{adv.FGSM} and determine that the large curvature along \emph{FGSM} perturbed direction leads to a large difference in perturbed directions generated by \emph{FGSM} and \emph{PGD} attacks, which account for the difference in robustness performances between \emph{adv.FGSM} and \emph{adv.PGD} (Figure~\ref{fig:1}). To deal with this we propose a regularization term that makes \emph{FGSM} perturbed direction close to \emph{PGD} perturbed direction, and allows for \emph{adv.FGSM} to reach comparable robustness performance as \emph{adv.PGD}. Our experimental studies demonstrate that the proposed method achieves comparable results on MNIST dataset and better results on CIFAR-10 dataset compared with \emph{adv.PGD}.

Our contributions are summarized as follows:
\begin{itemize}
    \item We analyze the influence of the curvature along \emph{FGSM} perturbed direction on the perturbations generated by \emph{FGSM} and \emph{PGD} attacks respectively. We show that the curvature along \emph{FGSM} perturbed direction has a significant influence on the performance of adversarial robustness achieved by \emph{adv.FGSM}.
    \item We develop a curvature regularization term for restraining the curvature along \emph{FGSM} perturbed direction when training model with \emph{adv.FGSM}, which is called as \emph{adv.FGSMR} method. \emph{adv.FGSMR} can effectively bridge the performance gap between \emph{adv.FGSM} and \emph{adv.PGD}.
    \item Extensive experiments show that \emph{adv.FGSMR} achieves comparable performance on MNIST under white-box attack. Besides, it achieves better performance on CIFAR-10 under white-box attack and effectively defends the transferable adversarial attack as well. Experiments also show that \emph{adv.FGSMR} achieves comparable convergence speed on perturbed-data accuracy during training process while requires only half of the time for training one epoch compared with \emph{adv.PGD}.
\end{itemize}

The rest of this paper is organized as follows. Section~\ref{prelim} describes the preliminary knowledge. Section~\ref{method} presents the proposed method for bridging the performance gap between \emph{adv.FGSM} and \emph{adv.PGD}. Section~\ref{experiments} introduces evaluations in terms of training efficiency and adversarial robustness, and Section~\ref{discuss} discusses reasons for the better performance of our proposed \emph{adv.FGSMR} on CIFAR-10 dataset than \emph{adv.PGD}. Finally, Section~\ref{conclusion} draws the conclusions of this study. 

\section{Preliminaries}\label{prelim}
\subsection{Notation}
We denote our deep neural network as $f_{\theta}(x)$ where $x \in R^{d}$ is an instance of input data, and $L(f_{\theta}(x),y)$ is the \emph{cross-entropy} loss where $y$ is the true label. $sgn$ denotes the sign function. $\nabla_{x}L(\cdot)$ denotes the gradient of $L(\cdot)$ with respect to $x$. $S$ is the set constrained by $l_\infty$ or $l_2$ ball. $\epsilon$ is the allowed perturbation size. $k$ is the total iterations for \emph{PGD} attacks. 
\subsection{Adversarial Attacks}
Recently, various powerful adversarial attacks have been proposed to change models' prediction by adding small carefully designed perturbations. Several state-of-the-art  methods (i.e.~\emph{PGD}~\cite{Madry2017}, \emph{FGSM}~\cite{Goodfellow2014}, \emph{Deepfool}~\cite{Moosavi-Dezfooli2016}, \emph{C\&W}~\cite{carlini2017towards}) will be used to test the performance of defense models in this paper, which will be briefly introduced as follows.

\textbf{Fast Gradient Sign Method (\emph{FGSM})}~\cite{Goodfellow2014} obtains adversarial examples by the following equation:
\begin{align}
x^{*}=x+\epsilon\cdot sgn(\nabla_x L(f_{\theta}(x),y)).
\label{eq:1}
\end{align}

\textbf{Projected Gradient Descent (\emph{PGD})}~\cite{Madry2017} obtains adversarial examples by multi-step variant \emph{FGSM}. With the initialization $x^{0}=x$, the perturbed data in $t$-th step $x^{t}$ can be expressed as follows:
\begin{equation}
x^{t}=\Pi_{x+S}(x^{t-1}+\alpha\cdot sgn(\nabla_x L(f_{\theta}(x^{t-1}),y))) ,
\label{eq:2}
\end{equation}
where $\Pi_{x+S}$ denotes projecting perturbations into the set $S$ and $\alpha$ is the step size. We denote \emph{PGD} bounded with $l_\infty$ as \emph{PGD-inf} attack, and \emph{PGD} bounded with $l_2$  as \emph{PGD-l2} attack.
 
\textbf{Deepfool}~\cite{Moosavi-Dezfooli2016} computes adversarial perturbation of minimal norm for an given input in an iterative way. It finds the nearest decision boundary for generating perturbations by multiple linearization of the classifier. 

\textbf{\emph{C\&W} Attack}~\cite{carlini2017towards} generates adversarial examples by optimizing the $l_p$-norm of distance of $\delta$ with respect to the given input data $x$, which can be described as:
\begin{align}
    \min_{\delta} \;\; \norm{\delta}_p+c\cdot L(x+\delta)  \;\;  s.t. \;\;    x+\delta\in [0,1]^n ,
\label{eq:3}
\end{align}
where $\delta$ is the optimized perturbation for input $x$ and $c$ is a constant.

\textbf{Black-box Attack}. A popular kind of black-box attack utilizes cross-model transferability of adversarial samples~\cite{papernot2017practical,liu2016delving}, which trains a local substitute model to generate adversarial examples and tests it on another model. In this work, we specifically carry out the transferable adversarial attack~\cite{liu2016delving} as black-box attack evaluation. 

\subsection{Adversarial Training} Different from  \emph{Vanilla training}, adversarial training uses adversarial samples instead of clean samples to train model. Generally, the optimization function of adversarial training can be represented as follows~\cite{Madry2017}: 
\begin{align}
 \min_{\theta}\rho(\theta),  \rho(\theta)=\mathbbm{E}_{(x,y)\sim D}[\max_{\delta \in S} L(f_{\theta}(x+\delta),y)] .
 \label{eq:4}
\end{align}
In this paper, we call it \emph{adv.PGD} method when $\max_{\delta \in S} L(f_{\theta}(x+\delta),y)$ is solved by \emph{PGD-inf} attack. Similarly, we call it \emph{adv.FGSM} method when $\max_{\delta \in S} L(f_{\theta}(x+\delta),y)$ is solved by \emph{FGSM} attack. It is easy to see that \emph{adv.PGD} takes much more training time than \emph{adv.FGSM} since \emph{PGD-inf} attack takes multiple iterations while \emph{FGSM} attack take only one iteration.

\section{Methodology}\label{method}
In this section, we first give a fully analysis to explain why \emph{adv.FGSM} can not achieve the performance of adversarial robustness with \emph{adv.PGD}. Based on the analysis, we further extend \emph{adv.FGSM} in order to achieve comparable performance with \emph{adv.PGD}.

\subsection{Analysis of Performance Gap between adv.FGSM and adv.PGD}\label{analysis}
Considering the only difference between \emph{adv.FGSM} and \emph{adv.PGD} is that the adversarial examples are generated by \emph{FGSM} attack or \emph{PGD-inf} attack. Thus we first mainly explore the perturbation difference generated by \emph{FGSM} and \emph{PGD-inf} respectively. From the definitions of \emph{PGD-inf} and \emph{FGSM} attacks in Section~\ref{prelim}, we know \emph{PGD-inf} attack is a multi-step variant of \emph{FGSM} attack and it is apparent that \emph{PGD-inf} attack can generate more accurate perturbation compared with \emph{FGSM} attack. Figure~\ref{fig:2} shows the simplified schematic of \emph{PGD-inf} and \emph{FGSM} attacks. It indicates that the difference of perturbed directions generated by them will be enlarged with the increasing of the curvature along \emph{FGSM} perturbed direction. We believe that a large difference in perturbed directions will lead to the radical difference in adversarial robustness performance achieved by \emph{adv.FGSM} and \emph{adv.PGD} because the perturbed training dataset depends on these perturbed directions. This conjecture is supported by the experiment in Figure~\ref{fig:3a}. Therefore, we propose that as long as the curvature along \emph{FGSM} perturbed direction is kept to be small during training process, \emph{adv.FGSM} can achieve comparable performance with \emph{adv.PGD} for the following reasons:
\begin{figure}
    \centering
        \includegraphics[width=0.46\textwidth]{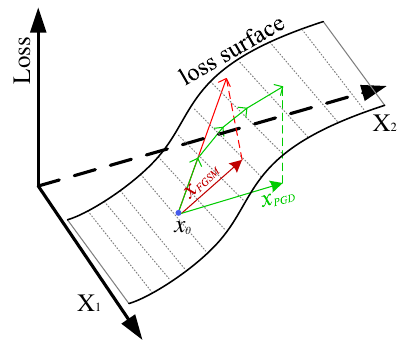}
    \caption{The simplified schematic diagram for perturbed directions generated by \emph{PGD-inf} and \emph{FGSM} attacks. Red arrow $\textbf{a}$ shows the perturbed direction of \emph{FGSM} attack and green arrows $\textbf{b}$ and $\textbf{c}$ show the perturbed direction of the two-step \emph{PGD-inf} attack. $x_0$ is a specific input. Due to the curvature, the perturbed directions generated by \emph{PGD-inf} and \emph{FGSM} can't be identical.}
    \label{fig:2}
\end{figure}
\begin{itemize}
    \item The perturbed directions generated by \emph{FGSM} and \emph{PGD-inf} attacks will be approaching to be identical with the curvature along \emph{FGSM} perturbed direction approaching to zero (Figure~\ref{fig:2}). As soon as the perturbed directions are the same, the perturbed training set will also be the same since the size of perturbation has the same constraint, and consequently the performance of adversarial robustness between \emph{adv.FGSM} and \emph{adv.PGD} should be the same.
    \item During \emph{adv.FGSM} training process, the perturbed-data accuracy under \emph{PGD-inf} attack stops increasing until the curvature along \emph{FGSM} perturbed direction surges suddenly (Figure~\ref{fig:3a}). This provides evidence that the curvature along \emph{FGSM} has a significant influence on the adversarial robustness performance of \emph{adv.FGSM}.
    \item The curvature along \emph{FGSM} perturbed direction is also kept to be small during \emph{adv.PGD} training process (Figure~\ref{fig:3b}). It indicates that to restrain the growth of the curvature value is reasonable.
\end{itemize} 
\begin{figure*}[htb]
    \centering
    \subfloat[]{
        \includegraphics[width=0.46\textwidth]{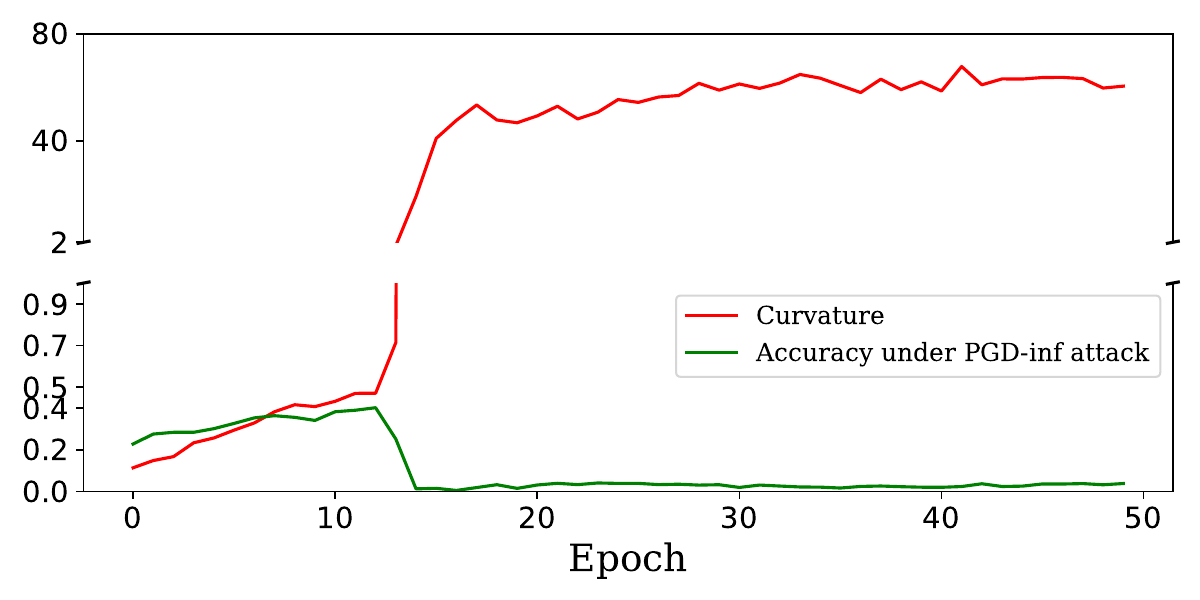}
        \label{fig:3a}
    }
    \subfloat[]{
        \includegraphics[width=0.46\textwidth]{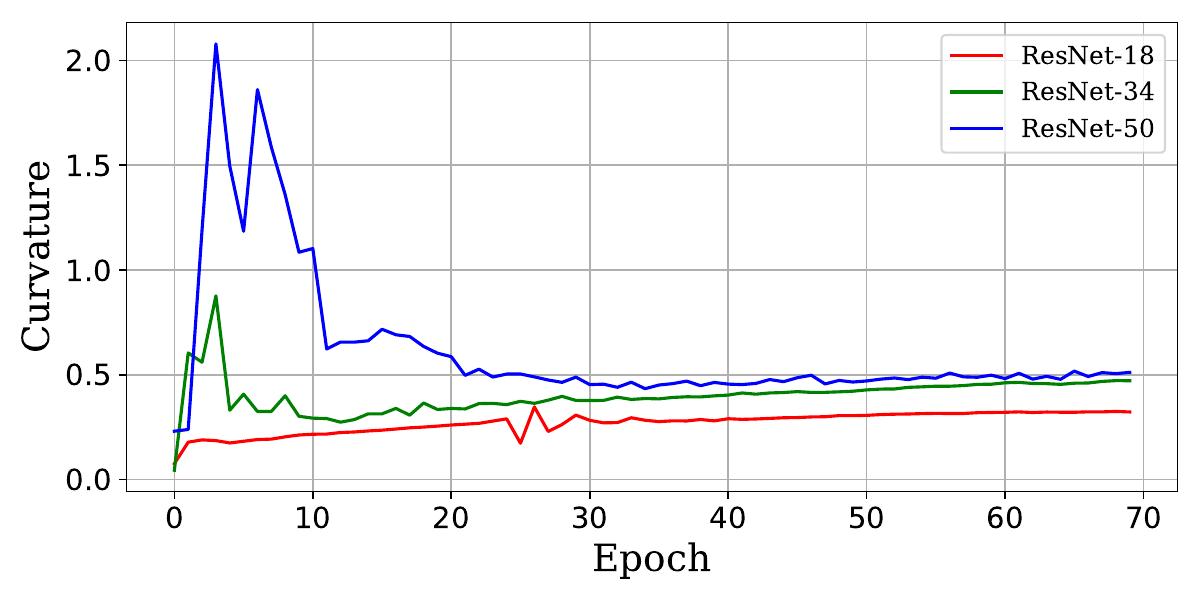}
        \label{fig:3b}
    }
    \caption{ \emph{(a)}: The average curvature along \emph{FGSM} perturbed direction on CIFAR-10 training set and the perturbed-data accuracy curve on perturbed test set generated by \emph{PGD-inf} attack. The training process is based on \emph{ResNet-18} model and \emph{adv.FGSM}. \emph{(b)}: The average curvature along \emph{FGSM} perturbed direction on CIFAR-10 training set during the training process with \emph{adv.PGD}. The curvature value is calculated using Eq.~\ref{eq6}.}
    \label{fig:3}
\end{figure*}
\subsection{Proposed Method}
Based on the descriptions in Section~\ref{analysis}, we propose to use a curvature regularization for restraining the growth of the curvature value and making \emph{FGSM} perturbed direction close to \emph{PGD-inf} perturbed direction. Formally, Let $L_{\theta}(x)$ be the cross-entropy loss; $g=sgn(\nabla_{x}L_{\theta}(x))$ be the \emph{FGSM} perturbed direction at data point $x$; $\delta=\epsilon g$ be the perturbation generated by \emph{FGSM} attack. As what we want to restrain is the gradient variation along \emph{FGSM} perturbed direction, namely, the second directional derivative, here we want to emphasize that the curvature value corresponds to the second directional derivative instead of the exact definition of curvature. According to the definition of  the directional derivative, the second derivative along \emph{FGSM} perturbed direction can be represented as:
\begin{align}
    \nabla^2_{xg}L_{\theta}(x)=\lim_{h  \to 0} \frac{\nabla_{x}L_{\theta}(x+hg)-\nabla_{x}L_{\theta}(x)}{h} .
    \label{eq:5}
\end{align}
Following the paper~\cite{Moosavi-Dezfooli2018}, by using a finite difference approximation, we have $\nabla^2_{xg}L_{\theta}(x)=\frac{\nabla_{x}L_{\theta}(x+hg)-\nabla_{x}L_{\theta}(x)}{h}$. The denominator can be omitted since it is a constant. Therefore, we give the curvature regularization term as follows:
\begin{align}
    R_{\theta}=\norm{\nabla_{x}L_{\theta}(x+hg)-\nabla_{x}L_{\theta}(x)}_2    ,
    \label{eq6}
\end{align}
where $h$ is set to $\epsilon$. The form of Eq.~\ref{eq6} is similar to \emph{CURE} method~\cite{Moosavi-Dezfooli2018} but a difference is that the perturbation size here is fixed and the perturbed direction is generated by \emph{FGSM} attack. 
The adversarial training optimization goal is to minimize the following expression:
\begin{align}
    \min_{\theta} L_{\theta}(x+\epsilon g)+\lambda R_{\theta}    ,
    \label{eq7}
\end{align}
where $R_{\theta}$ is the curvature regularization defined in Eq.~\ref{eq6}. $\lambda$ is the hyperparameter to control the strength of penalizing the curvature along \emph{FGSM} perturbed direction.

\section{Experiments}\label{experiments}
\subsection{Experiments Setup}
\paragraph{Datasets and network architectures} All experiments are run on MNIST and CIFAR-10 datasets. MNIST~\cite{lecun1998gradient} consists of 28x28 gray-scale images for handwritten digits with 60K training images and 10K test images. CIFAR-10~\cite{krizhevsky2009learning} consists of 32x32 color images that contain 10 different classes with 50K training images and 10K test images.

For MNIST dataset, we use a simple convolutional neural network with four convolution and two dense layers as our model architecture. For CIFAR-10 dataset, the Residual Networks-18/34/50~\cite{He2016} and Wide Residual Networks-$22\times 1/5/10\times 0\times10$~\cite{zagoruyko2016wide} are used as our model architecture. For comparison, robust models trained by \emph{adv.PGD} and \emph{adv.FGSM} respectively are evaluated as well. Please refer to the supplementary material for detailed training process.
\paragraph{Adversarial attacks}
In order to have a comprehensive evaluation for model's robustness, state-of-the-art white-box attacks are employed here. In specific, \emph{PGD-inf}, \emph{PGD-l2}, \emph{FGSM}, \emph{C\&W-l2} and \emph{Deepfool-l2} are used for white-box attack. By default, the hyperparameter $k$ is set to $20$ for \emph{PGD-inf/l2} in this paper. The accuracy on perturbed test set is used as adversarial robustness indicator, but for \emph{C\&W-l2} and \emph{Deepfool-l2} attacks, as they can find the adversarial examples that change the model's prediction for all inputs, we use the distance of the perturbed example to the clean example as the adversarial robustness evaluation indicator, refer to~\cite{Moosavi-Dezfooli2016}, the average distances is defined as follows:
\begin{align}
    \rho_{adv}=\frac{1}{\abs{\mathcal{D}}}\sum\limits_{x \in \mathcal{D}} \frac{\norm{x_{adv}-x}_2}{\norm{x}_2}  ,
    \label{eq8}
\end{align}
where $x_{adv}$ is the adversarial example generated by the attack algorithm, and $\mathcal{D}$ is the test set. \emph{C\&W-l2} and \emph{Deepfool-l2} attack are carried out by public attack tool:~\emph{foolbox}~\cite{rauber2017foolbox} and parameters are set by default for these two attacks. Beyond white-box attack, the transferable adversarial attack~\cite{liu2016delving} is employed on CIFAR-10 dataset as block-box attack evaluation.

\subsection{Training Efficiency}
\begin{figure}[htb]
    \centering
    \includegraphics[width=0.5\textwidth]{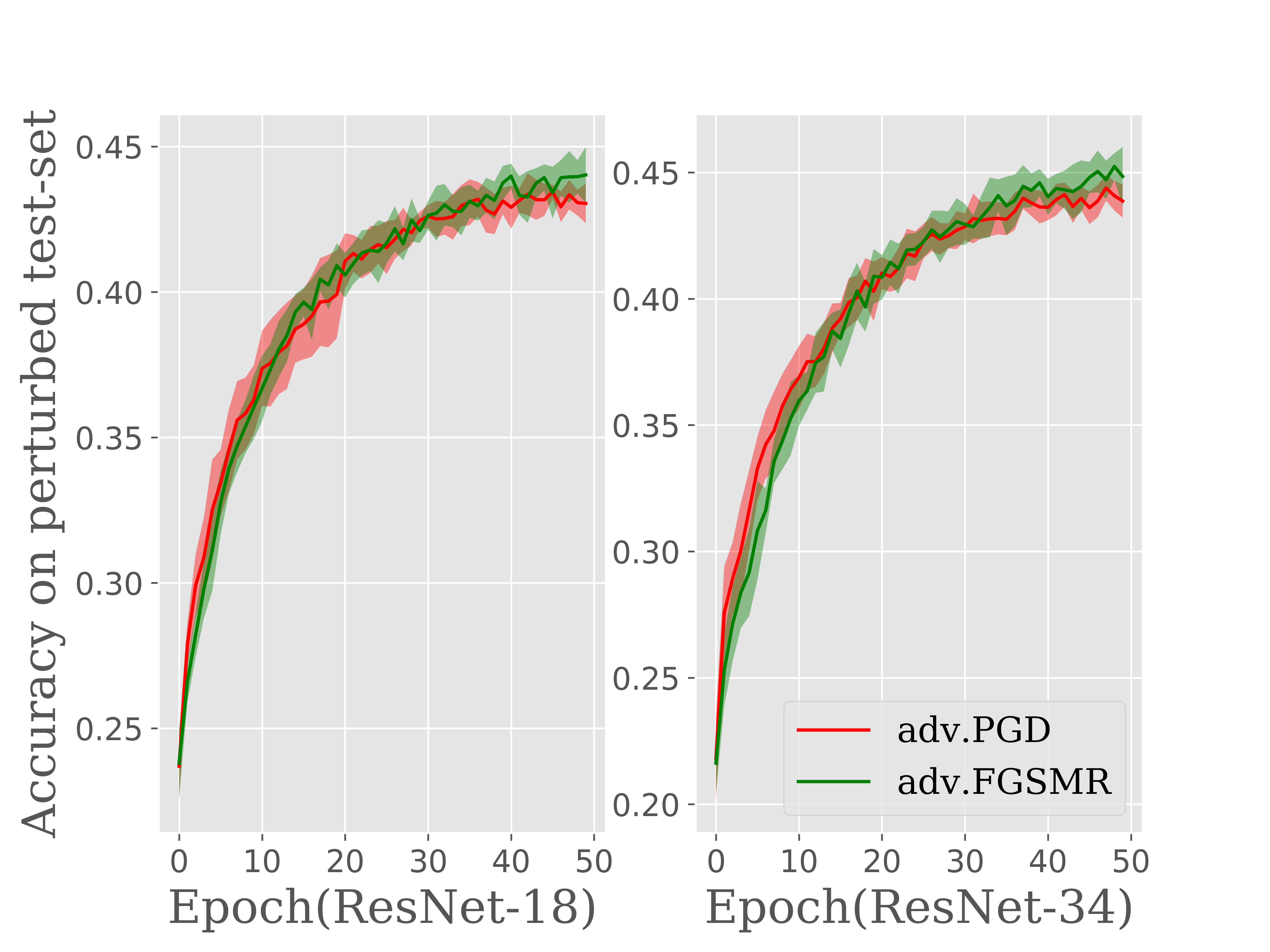}
    \caption{A comparable convergence speed on perturbed-data accuracy between \emph{adv.FGSMR} and \emph{adv.PGD}. Left figure: the training process of \emph{ResNet-18} model. Right figure: the training process of \emph{ResNet-34} model. Perturbed test set are generated by \emph{PGD-inf} attack ($\epsilon=8.0/255$) on CIFAR-10 test set. The accuracy variation for each epoch is plotted using one standard deviation.}
    \label{fig3}
\end{figure}

We evaluate the training efficiency of \emph{adv.FGSMR} and compare it with \emph{adv.PGD}. The training efficiency is evaluated from two aspects: (1) how much time does it take for training one epoch; and (2) how fast can the adversarial robustness be improved during training process. For the first aspect, as \emph{adv.PGD} method uses \emph{PGD-inf} attack to generate perturbed examples, it takes $k$ (commonly $k$ is set to 20) times of forward and backward process where $k$ is the total iterations for \emph{PGD-inf} attack. But for \emph{adv.FGSMR}, it takes $1$ time of forward and backward process to generate perturbed examples plus $2$ times of forward and backward process for the curvature regularization. Therefore, from the analysis above, \emph{adv.FGSMR} saves $(k-3)$ times of forward and backward process. Table~\ref{tb:traintime} shows the training time of 50 epochs for \emph{adv.PGD} ($k=20$) and \emph{adv.FGSMR} respectively, which indicates that \emph{adv.FGSMR} takes half time of what \emph{adv.PGD} ($k=20$) takes. For the second aspect, we record the perturbed-data accuracy under \emph{PGD-inf} attack on CIFAR-10 test set with first 50 training epochs for \emph{adv.PGD} and \emph{adv.FGSMR} with $\epsilon=8.0/255$ respectively. We repeat the training process 10 times and report the mean and standard deviation. The results (Figure~\ref{fig3}) show that the perturbed-data accuracy of \emph{adv.FGSMR} can be converged as fast as \emph{adv.PGD}. Therefore, we conclude that \emph{adv.FGSMR} has higher training efficiency since it takes less time for training one epoch and has comparable convergence speed upon the perturbed-data accuracy compared with \emph{adv.PGD}.  

\begin{table*}[!t]
\centering  
\begin{tabular}{ccccc} \\
\thickhline
Time  &ResNet-18 &ResNet-34 &ResNet-18 &ResNet-34\\
(minutes)  &(adv.PGD) &(adv.PGD)&(Ours) &(Ours) \\
\hline 
Training time(50 Epoch)   &214 &375&106   &187 \\
\thickhline
\end{tabular}
\caption{ Comparison of time spent on training 50 epochs with \emph{adv.PGD} and \emph{adv.FGSMR} respectively. This experiment is based on CIFAR-10 dataset.}
\label{tb:traintime}
\end{table*}

\subsection{Performance under White-box Attack}
\paragraph{Performance on MNIST Dataset}
We evaluate the performance of our proposed \emph{adv.FGSMR} on MNIST dataset. For comparison, the performance of \emph{adv.PGD}, \emph{adv.FGSM} and \emph{CURE}~\cite{Moosavi-Dezfooli2018} are shown. Robust models with $\epsilon=0.1$ and $\epsilon=0.2$ are trained by \emph{adv.PGD}, \emph{adv.FGSM} and \emph{adv.FGSMR} respectively.
Various state-of-the-art attacks are used for evaluating adversarial robustness including \emph{FGSM}, \emph{PGD-l2}, \emph{PGD-inf}, \emph{Deepfool-l2} and \emph{C\&W-l2} attacks. The  hyperparameter $\epsilon$ is set to $0.2$, $2$, $0.1$ for \emph{FGSM}, \emph{PGD-l2} and \emph{PGD-inf} attacks respectively.

From Table \ref{tab3}, We can see that our method achieves higher perturbed-data accuracy than \emph{adv.PGD} under \emph{FGSM}, \emph{PGD-l2} and \emph{PGD-inf} attacks. For \emph{Deepfool-l2} attack, the average distance $\rho_{adv}$ values of our method are slightly smaller than that of \emph{adv.PGD}. For \emph{C\&W-l2} attack, our method achieves slightly larger distance on robust model with $\epsilon=0.2$ while achieves slightly smaller distance on robust model with $\epsilon=0.1$. It is also worthy to note that our method achieves state-of-the-art accuracy on clean test set. In general, our method achieves comparable adversarial robustness performance compared with \emph{adv.PGD}.

For \emph{adv.FGSM}, it achieves better performance on \emph{FGSM} attack but performs worse on the other four attacks, which is consistent with the results reported in~\cite{kurakin2016adversarial}. Considering the curvature regularization is similar to \emph{CURE} method~\cite{Moosavi-Dezfooli2018}, we also show the performance of \emph{CURE} method that is proposed to improve robustness by decreasing the curvature of loss function. The results (Table~\ref{tab3}) show that the performance achieved by \emph{CURE} is obviously worse than the performance achieved by \emph{adv.PGD} and \emph{adv.FGSMR} under all attacks.

\begin{table*}[htb]
\centering  
\begin{tabular}{lcccccc} \\
\thickhline
\multicolumn{1}{c}{\multirow{2}{*}{\diagbox{Training methods}{Attack methods}}}  &Clean &\emph{FGSM} &\emph{PGD-l2} &\emph{PGD-inf} &\emph{Deepfool-l2} &\emph{C\&W-l2}\\
 &(accuracy) &(accuracy) &(accuracy) &(accuracy) &($\rho_{adv}$) &($\rho_{adv}$)\\ 
\hline 
\emph{Vanilla train}                &0.98  &0.361  &0.448 &0.27 &0.54 &0.46  \\
\emph{adv.PGD}($\epsilon:0.1$) &0.993 &0.897 &0.956 &0.974 &1.25 &0.85 \\
\emph{adv.PGD($\epsilon:0.2$)} &0.992 &0.966      &0.975 & 0.982      &1.36    & 0.87     \\
\emph{adv.FGSM($\epsilon:0.1$)} &0.992&0.988 &0.950 &       0.971&1.02    &0.77       \\
\emph{adv.FGSM($\epsilon:0.2$)} &0.993     &0.968      &0.950 &       0.972&1.07 &0.66       \\
\emph{CURE~\cite{Moosavi-Dezfooli2018}}&0.990 &0.936 &0.932 &0.957 &1.02 &0.79 \\
\emph{\textbf{adv.FGSMR}}($\epsilon:0.1$) &0.994 &0.961 &0.959 &0.979 &1.15 &0.84 \\
\emph{\textbf{adv.FGSMR}}($\epsilon:0.2$) &0.992 &0.968 &0.976 &0.983 &1.31 &0.90\\
\thickhline
\end{tabular}
\caption{Performance of models trained by \emph{Vanilla train}, \emph{adv.PGD}, \emph{CURE}, \emph{adv.FGSMR} methods respectively against various attacks on MNIST Dataset. For \emph{FGSM}, \emph{PGD-l2} and \emph{PGD-inf} attacks, the accuracy on perturbed MNIST test set is taken as evaluation indicator. For \emph{Deepfool-l2} and \emph{C\&W-l2} attacks, the average distance ($\rho_{adv}$) is taken as the evaluation indicator and is calculated using Eq.~\ref{eq8}.}
\label{tab3}
\end{table*}

\paragraph{Performance on CIFAR-10 Dataset}
We show the adversarial robustness performance of the proposed \emph{adv.FGSMR} on CIFAR-10 dataset. For comparison, the adversarial robustness performance of \emph{adv.PGD} and \emph{Vanilla train} are also evaluated. For \emph{adv.FGSMR}, we train three robustness models with $\epsilon=8.0/255,9.0/255,10.0/255$ respectively. The same as on MNIST dataset, \emph{FGSM}, \emph{PGD-l2}, \emph{PGD-inf}, \emph{Deepfool-l2} and \emph{C\&W-l2} attacks are chosen for testing the adversarial robustness performance. The hyperparameter $\epsilon$ is set to $8.0/255$ for \emph{FGSM} and \emph{PGD-inf} attacks, and $60.0/255$ for \emph{PGD-l2} attack. 

The results (Table~\ref{tab4}) show that our method achieves higher perturbed-data accuracy than \emph{adv.train-PGD} under \emph{FGSM}, \emph{PGD-inf} and \emph{PGD-l2} attacks, and the average distance $\rho_{adv}$ values are larger than that of \emph{adv.PGD} under \emph{Deepfool-l2} and \emph{C\&W-l2} attacks. The large average distance $\rho_{adv}$ values indicate that our method indeed enlarges the distance of input $x$ to its nearest boundary. For \emph{adv.FGSM}, it achieves much higher accuracy on \emph{FGSM} perturbed examples than on clean examples, which is claimed as label leaking problem in~\cite{kurakin2016adversarial}. The average distance $\rho_{adv}$ also shows that the model trained by \emph{adv.FGSM} nearly does not enlarge the distance of input $x$ to the nearest decision boundary.

We also observe that with increasing perturbation size $\epsilon$ from $8.0/255$ to $10.0/255$, the clean accuracy decreases gradually and the perturbed-data accuracy under \emph{PGD-inf} attack increases gradually, which is consistent with the claim~\cite{Tsipras2018} that there is a trade-off between clean accuracy and adversarial robustness. However, it is interesting that the perturbed-data accuracy under \emph{FGSM} and \emph{PGD-l2} attacks does not show an increasing trend. We argue the perturbed-data accuracy might depend more on clean accuracy since the \emph{FGSM} and \emph{PGD-l2} attacks are weaker than \emph{PGD-inf} attack.  

\begin{table*}[htb]
\centering  
\begin{tabular}{lcccccc} \\
\thickhline
\multicolumn{1}{c}{\multirow{2}{*}{\diagbox{Training methods}{Attack methods}}}  &Clean &\emph{FGSM} &\emph{PGD-l2} &\emph{PGD-inf} &\emph{Deepfool-l2} &\emph{C\&W-l2}\\
&(accuracy) &(accuracy) &(accuracy) &(accuracy) &($\rho_{adv}$) &($\rho_{adv}$)\\ 
\hline 
\emph{Vanilla train} &0.909 &0.237  &0.308 &0.000 &0.031 &0.025  \\
\emph{adv.FGSM}($\epsilon:8.0/255$) &0.849 &0.908 &0.353 &0.091 & 0.022& 0.016\\
\emph{adv.PGD}($\epsilon:8.0/255$) &0.746 &0.506 &0.710 &0.444 &0.178 &0.129\\
\emph{\textbf{adv.FGSMR}}($\epsilon:8.0/255$) &0.789 &0.51 &0.759 &0.458 &0.228 &0.179 \\
\emph{\textbf{adv.FGSMR}}($\epsilon:9.0/255$) &0.772 &0.507 &0.734 &0.465 &0.227 &0.180\\
\emph{\textbf{adv.FGSMR}}($\epsilon:10.0/255$) &0.756 &0.509 &0.723& 0.470 &0.230 &0.177\\
\thickhline
\end{tabular}
\caption{Performance of models trained by \emph{Vanilla train}, \emph{adv.train-PGD}, \emph{adv.train-FGSMR} methods respectively under various attacks on CIFAR-10 dataset. For \emph{FGSM} and \emph{PGD-inf/l2} attacks, the accuracy on perturbed CIFAR-10 test set is taken as evaluation indicator. For \emph{Deepfool-l2} and \emph{C\&W-l2} attacks, the average distance ($\rho_{adv}$) is taken as evaluation indicator. }
\label{tab4}
\end{table*}
\paragraph{Effect of network capacity}
In order to explore the relation between network capacity and adversarial robustness improved by \emph{adv.FGSMR} ($\epsilon=8.0/255$), we evaluate the adversarial robustness performance on \emph{ResNet-18/34/50} and \emph{Wide ResNet-22$\times$1/5/10$\times$0$\times10$} for different depths and widths respectively. Madry~\cite{Madry2017} concludes by experiments that increasing capacity of model can increase the model's adversarial robustness. In our results (Table \ref{tab5}), the perturbed-data accuracy achieved by \emph{adv.FGSMR} shows the same increasing tendency both with the increasing of the model's width or depth, which is consistent with the claim of~\cite{Madry2017}. Besides, the perturbed-data accuracy achieved by our method is all higher than the perturbed-data accuracy achieved by \emph{adv.PGD}, which further provides evidences that the proposed method achieves better performance on CIFAR-10 dataset. We also calculate the average curvature for the six models where the average curvature is calculated using Eq.~\ref{eq6}. The results (Table~\ref{tab5}) show the curvature values are smaller than the curvature values of \emph{adv.PGD}, which indicates the curvature value can be effectively restrained by our proposed regularization. 
\begin{table*}[htb]
\centering
\begin{tabular}{lccccccc}\\
\thickhline
     &Capacity  &\multicolumn{3}{c}{\emph{adv.PGD}} &\multicolumn{3}{c}{\emph{\textbf{adv.FGSMR}}} \\
\cline{3-8}
    Models&(Million)  &\emph{PGD-inf} &\emph{FGSM} & Average Curvature&\emph{PGD-inf} &\emph{FGSM} &Average Curvature \\
    
\hline
    \emph{ResNet-18}&11  &0.444 &0.506 &0.487 &0.458 &0.51 &0.324 \\
    \emph{ResNet-34}&21  &0.469 &0.511 &0.442 &0.475& 0.525 &0.309 \\
    \emph{ResNet-50}&23  &0.448 &0.512 &0.565 &0.479 &0.528 &0.338\\
\hline
    \emph{WResNet-22x1}&0.27 &0.383 &0.407 &0.282 &0.408 &0.438 &0.245\\
    \emph{WResNet-22x5}&6   &0.438 &0.495 &0.502 & 0.462&0.495 &0.262\\
    \emph{WResNet-22x10}&26  &0.440 &0.498 &0.504 & 0.477&0.515 &0.319\\
\thickhline
\end{tabular}
\caption{Effect of network depth and width. The perturbed-data accuracy under \emph{PGD-inf} and \emph{FGSM} attacks are shown for robust models with different capacity. For network depth, \emph{ResNet-18/34/50} with increasing depth are reported. For network width, \emph{Wide ResNet-22$\times$1/5/10$\times$0$\times$10} with increasing width are reported. As comparing, Robust models achieved by \emph{adv.PGD} are tested too. Capacity denotes the number of trainable parameters in the model.}
\label{tab5}
\end{table*}
\subsection{Performance under Black-box Attack}
In this section, we evaluate our proposed method based on transferable adversarial attack~\cite{liu2016delving}. Following the transferable adversarial attack, three no-defense models and two robust models are trained for source model and six models trained by \emph{Vanilla train}, \emph{adv.FGSMR} ($\epsilon=8.0/255$) and \emph{adv.PGD} ($\epsilon=8.0/255$) methods respectively are taken as target model. The adversarial examples under \emph{PGD-inf} attack with ($\epsilon=8.0/255$) are generated from source model to attack target model. The results (Table~\ref{tab6}) show that models trained by \emph{adv.FGSMR} achieve slightly higher perturbed-data accuracy than models trained by \emph{adv.PGD} under transferable adversarial examples generated from both robust and non-defense models, which indicates \emph{adv.FGSMR} can defend black-box attack as effective as \emph{adv.PGD}. We also observe that the perturbed-data accuracy achieved by \emph{adv.FGSMR} are much more close to \emph{adv.PGD} than \emph{Vanilla train}, which indicates that \emph{adv.FGSMR} learns a similar feature with \emph{adv.PGD} but a different feature with \emph{Vanilla train}.

\begin{table*}[htb]
\centering
\begin{tabular}{cccccccc}\\
\thickhline
    \multicolumn{2}{c}{\multirow{2}{*}{\diagbox{Source model}{Target model}}} &\multicolumn{2}{c}{\emph{Vanilla train}}&\multicolumn{2}{c}{\emph{adv.PGD}}&\multicolumn{2}{c}{\emph{\textbf{adv.FGSMR}}}\\
\cline{3-8}
    \multicolumn{2}{c}{\multirow{2}{*}{}}&\emph{ResNet-18}&\emph{ResNet-34}&\emph{ResNet-18} &\emph{ResNet-34} &\emph{ResNet-18} &\emph{ResNet-34}\\
\hline
    \multicolumn{2}{l}{\emph{Vanilla train}(\emph{ResNet-18})}&0.00&0.040 &0.736 &0.759&0.771 &0.764 \\
    \multicolumn{2}{l}{\emph{Vanilla train}(\emph{ResNet-34})}&0.070&0.016 &0.735 &0.758&0.772 &0.764 \\
    \multicolumn{2}{l}{\emph{Vanilla train}(\emph{ResNet-50})}&0.071&0.084 & 0.747&0.760&0.774 &0.766 \\
    \multicolumn{2}{l}{\emph{adv.PGD}(\emph{ResNet-18})}&0.792&0.787  &0.444&0.584&0.606 &0.614 \\
    \multicolumn{2}{l}{\emph{adv.PGD}(\emph{ResNet-34})}&0.738&0.741 &0.554 &0.469&0.577 &0.582 \\
\thickhline
\end{tabular}
\caption{Against black-box attack. This table shows the perturbed-data accuracy under transferable adversarial attack. The rows denotes the three vanilla trained models and two robust models which are used for generating transferable adversarial examples on CIFAR-10 test set. The columns denotes models trained by \emph{Vanilla train}, \emph{adv.FGSMR} and \emph{adv.PGD} respectively that are used for testing.}
\label{tab6}
\end{table*}
\section{Discussion}\label{discuss}
In this paper, we first analyze the difference in \emph{FGSM} and \emph{PGD-inf} attacks and conclude that decreasing the curvature along \emph{FGSM} perturbed direction can increase the similarity between the perturbed directions generated by \emph{PGD-inf} and \emph{FGSM} attacks. Therefore, we use an extra curvature regularization to restrain the growth of the curvature in order to make \emph{FGSM} perturbed direction close to \emph{PGD-inf} perturbed direction. In our expectation, \emph{adv.FGSMR} can achieve the performance of \emph{adv.PGD}, however, in our experiments, the model trained by \emph{adv.FGSMR} achieves better adversarial robustness than the model trained by \emph{adv.PGD} on CIFAR-10 dataset. 
In order to provide the possible explanations for this behavior, we analyze the differences between \emph{adv.FGSMR} and \emph{adv.PGD}.

Firstly, \emph{adv.FGSMR} uses an extra regularization to control the curvature value while \emph{adv.PGD} does not. We observe from Table~\ref{tab6} that the curvature value under \emph{adv.FGSMR} is slightly smaller than the curvature value under \emph{adv.PGD}. We think that the smaller curvature might account for the better performance of \emph{adv.FGSMR} because it has been reported in~\cite{Moosavi-Dezfooli2018} that decreasing the curvature can improve the upper bound of the distance of input $x$ to its nearest decision boundary, namely the adversarial robustness. We also observe that the adversarial robustness improved by only decreasing the curvature of loss function does not achieve the performance of \emph{adv.PGD}~\cite{Moosavi-Dezfooli2018}, which indicates that methods based on adversarial training might be better in improving the lower bound of the adversarial robustness.
\section{Conclusion}\label{conclusion}
In this paper, we bridge the performance gap between \emph{adv.FGSM} and \emph{adv.PGD} methods by adding a curvature regularization. Firstly, we explore the reasons why \emph{adv.FGSM} can not achieve comparable performance with the \emph{adv.PGD}. We show that the difference of perturbed directions generated by \emph{PGD-inf} and \emph{FGSM} attacks respectively will increase with the increasing of the curvature along \emph{FGSM} perturbed direction. The large difference in perturbed directions finally leads to a large difference in performance on adversarial robustness. Based on this analysis, we propose that adding a curvature regularization to restrain the growth of curvature along \emph{FGSM} perturbed direction when training model with \emph{adv.FGSM}. We evaluate the proposed \emph{adv.FGSMR} in terms of training efficiency and adversarial robustness. Experiments show that \emph{adv.FGSMR} achieves comparable convergence speed on perturbed-data accuracy during training process but only takes half time for training one epoch compared with \emph{adv.PGD} ($k=20$). 
Experiments also show that, under white-box attack, the \emph{adv.FGSMR} achieves comparable performance on MNIST dataset and achieves better performance on CIFAR-10 dataset than \emph{adv.PGD}, under black-box attack, the \emph{adv.FGSMR} can defend the transferable adversarial attack as effective as \emph{adv.PGD}. 
{\small
\bibliographystyle{main}
\bibliography{main}

\begin{thebibliography}{10}\itemsep=-1pt

\bibitem{Akhtar2018}
Naveed Akhtar and Ajmal Mian.
\newblock {Threat of Adversarial Attacks on Deep Learning in Computer Vision: A
  Survey}, 2018.

\bibitem{athalye2018obfuscated}
Anish Athalye, Nicholas Carlini, and David Wagner.
\newblock Obfuscated gradients give a false sense of security: Circumventing
  defenses to adversarial examples.
\newblock {\em arXiv preprint arXiv:1802.00420}, 2018.

\bibitem{carlini2017towards}
Nicholas Carlini and David Wagner.
\newblock Towards evaluating the robustness of neural networks.
\newblock In {\em 2017 IEEE Symposium on Security and Privacy (SP)}, pages
  39--57. IEEE, 2017.

\bibitem{chen2017zoo}
Pin-Yu Chen, Huan Zhang, Yash Sharma, Jinfeng Yi, and Cho-Jui Hsieh.
\newblock Zoo: Zeroth order optimization based black-box attacks to deep neural
  networks without training substitute models.
\newblock In {\em Proceedings of the 10th ACM Workshop on Artificial
  Intelligence and Security}, pages 15--26. ACM, 2017.

\bibitem{Girshick2014}
Ross Girshick, Jeff Donahue, Trevor Darrell, and Jitendra Malik.
\newblock {Rich feature hierarchies for accurate object detection and semantic
  segmentation}.
\newblock In {\em Proceedings of the IEEE Computer Society Conference on
  Computer Vision and Pattern Recognition}, 2014.

\bibitem{Goodfellow2014}
Ian~J. Goodfellow, Jonathon Shlens, and Christian Szegedy.
\newblock {Explaining and Harnessing Adversarial Examples}.
\newblock dec 2014.

\bibitem{He2016}
Kaiming He, Xiangyu Zhang, Shaoqing Ren, and Jian Sun.
\newblock {Deep residual learning for image recognition}.
\newblock In {\em Proceedings of the IEEE conference on computer vision and
  pattern recognition}, pages 770--778, 2016.

\bibitem{Hinton2012}
Geoffrey Hinton, Li Deng, Dong Yu, George~E Dahl, Abdel-rahman Mohamed, Navdeep
  Jaitly, Andrew Senior, Vincent Vanhoucke, Patrick Nguyen, Tara~N Sainath, and
  Brian Kingsbury.
\newblock {Deep Neural Networks for Acoustic Modeling in Speech Recognition}.
\newblock {\em Ieee Signal Processing Magazine}, 2012.

\bibitem{jakubovitz2018improving}
Daniel Jakubovitz and Raja Giryes.
\newblock Improving dnn robustness to adversarial attacks using jacobian
  regularization.
\newblock In {\em Proceedings of the European Conference on Computer Vision
  (ECCV)}, pages 514--529, 2018.

\bibitem{krizhevsky2009learning}
Alex Krizhevsky, Geoffrey Hinton, et~al.
\newblock Learning multiple layers of features from tiny images.
\newblock Technical report, Citeseer, 2009.

\bibitem{Krizhevsky2012}
Alex Krizhevsky and Geoffrey~E. Hinton.
\newblock {ImageNet Classification with Deep Convolutional Neural Networks}.
\newblock {\em Neural Information Processing Systems}, 2012.

\bibitem{kurakin2016adversarial}
Alexey Kurakin, Ian Goodfellow, and Samy Bengio.
\newblock Adversarial machine learning at scale.
\newblock {\em arXiv preprint arXiv:1611.01236}, 2016.

\bibitem{lecun1998gradient}
Yann LeCun, L{\'e}on Bottou, Yoshua Bengio, Patrick Haffner, et~al.
\newblock Gradient-based learning applied to document recognition.
\newblock {\em Proceedings of the IEEE}, 86(11):2278--2324, 1998.

\bibitem{liu2016delving}
Yanpei Liu, Xinyun Chen, Chang Liu, and Dawn Song.
\newblock Delving into transferable adversarial examples and black-box attacks.
\newblock {\em arXiv preprint arXiv:1611.02770}, 2016.

\bibitem{Long2015}
Jonathan Long, Evan Shelhamer, and Trevor Darrell.
\newblock {Fully Convolutional Networks for Semantic Segmentation ppt}.
\newblock In {\em CVPR 2015 Proceedings of the IEEE Conference on Computer
  Vision and Pattern Recognition}, 2015.

\bibitem{Madry2017}
Aleksander Madry, Aleksandar Makelov, Ludwig Schmidt, Dimitris Tsipras, and
  Adrian Vladu.
\newblock {Towards Deep Learning Models Resistant to Adversarial Attacks}.
\newblock jun 2017.

\bibitem{Moosavi-Dezfooli2016}
Seyed~Mohsen Moosavi-Dezfooli, Alhussein Fawzi, and Pascal Frossard.
\newblock {DeepFool: A Simple and Accurate Method to Fool Deep Neural
  Networks}.
\newblock In {\em Proceedings of the IEEE Computer Society Conference on
  Computer Vision and Pattern Recognition}, volume 2016-Decem, pages
  2574--2582. IEEE Computer Society, dec 2016.

\bibitem{Moosavi-Dezfooli2018}
Seyed-Mohsen Moosavi-Dezfooli, Alhussein Fawzi, Jonathan Uesato, and Pascal
  Frossard.
\newblock {Robustness via curvature regularization, and vice versa}.
\newblock nov 2018.

\bibitem{papernot2017practical}
Nicolas Papernot, Patrick McDaniel, Ian Goodfellow, Somesh Jha, Z~Berkay Celik,
  and Ananthram Swami.
\newblock Practical black-box attacks against machine learning.
\newblock In {\em Proceedings of the 2017 ACM on Asia conference on computer
  and communications security}, pages 506--519. ACM, 2017.

\bibitem{Papernot2016}
Nicolas Papernot, Patrick McDaniel, Xi Wu, Somesh Jha, and Ananthram Swami.
\newblock {Distillation as a Defense to Adversarial Perturbations Against Deep
  Neural Networks}.
\newblock In {\em Proceedings - 2016 IEEE Symposium on Security and Privacy, SP
  2016}, pages 582--597. Institute of Electrical and Electronics Engineers
  Inc., aug 2016.

\bibitem{Qiu2019}
Shilin Qiu, Qihe Liu, Shijie Zhou, and Chunjiang Wu.
\newblock {Review of artificial intelligence adversarial attack and defense
  technologies}, 2019.

\bibitem{rauber2017foolbox}
Jonas Rauber, Wieland Brendel, and Matthias Bethge.
\newblock Foolbox: A python toolbox to benchmark the robustness of machine
  learning models.
\newblock {\em arXiv preprint arXiv:1707.04131}, 2017.

\bibitem{Song2017}
Yang Song, Taesup Kim, Sebastian Nowozin, Stefano Ermon, and Nate Kushman.
\newblock {PixelDefend: Leveraging Generative Models to Understand and Defend
  against Adversarial Examples}.
\newblock oct 2017.

\bibitem{Szegedy2013}
Christian Szegedy, Wojciech Zaremba, Ilya Sutskever, Joan Bruna, Dumitru Erhan,
  Ian Goodfellow, and Rob Fergus.
\newblock {Intriguing properties of neural networks}.
\newblock dec 2013.

\bibitem{Tsipras2018}
Dimitris Tsipras, Shibani Santurkar, Logan Engstrom, Alexander Turner, and
  Aleksander Madry.
\newblock {Robustness May Be at Odds with Accuracy}.
\newblock may 2018.

\bibitem{wang2019convergence}
Yisen Wang, Xingjun Ma, James Bailey, Jinfeng Yi, Bowen Zhou, and Quanquan Gu.
\newblock On the convergence and robustness of adversarial training.
\newblock In {\em International Conference on Machine Learning}, pages
  6586--6595, 2019.

\bibitem{zagoruyko2016wide}
Sergey Zagoruyko and Nikos Komodakis.
\newblock Wide residual networks.
\newblock {\em arXiv preprint arXiv:1605.07146}, 2016.

\end{thebibliography}
}

\end{document}


\title{Supplementary materials for bridging the Performance Gap between FGSM and PGD Adversarial Training}
\maketitle
\section{Training Process}
\subsection{MNIST}
\begin{figure}[htb]
    \centering
     \minipage{0.35\textwidth}
        \includegraphics[width=\linewidth]{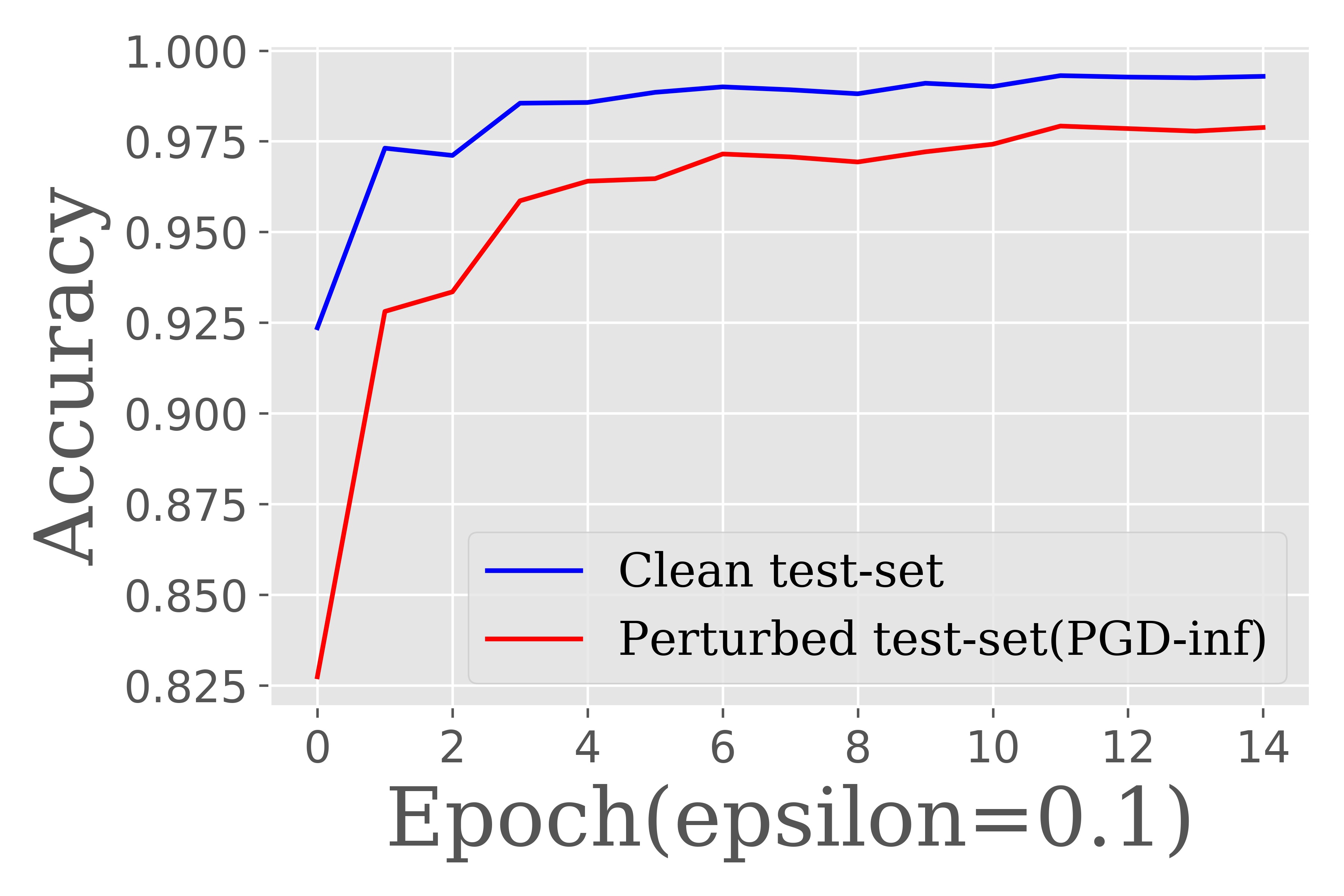}
    \endminipage\hfill
    \minipage{0.35\textwidth}
        \includegraphics[width=\linewidth]{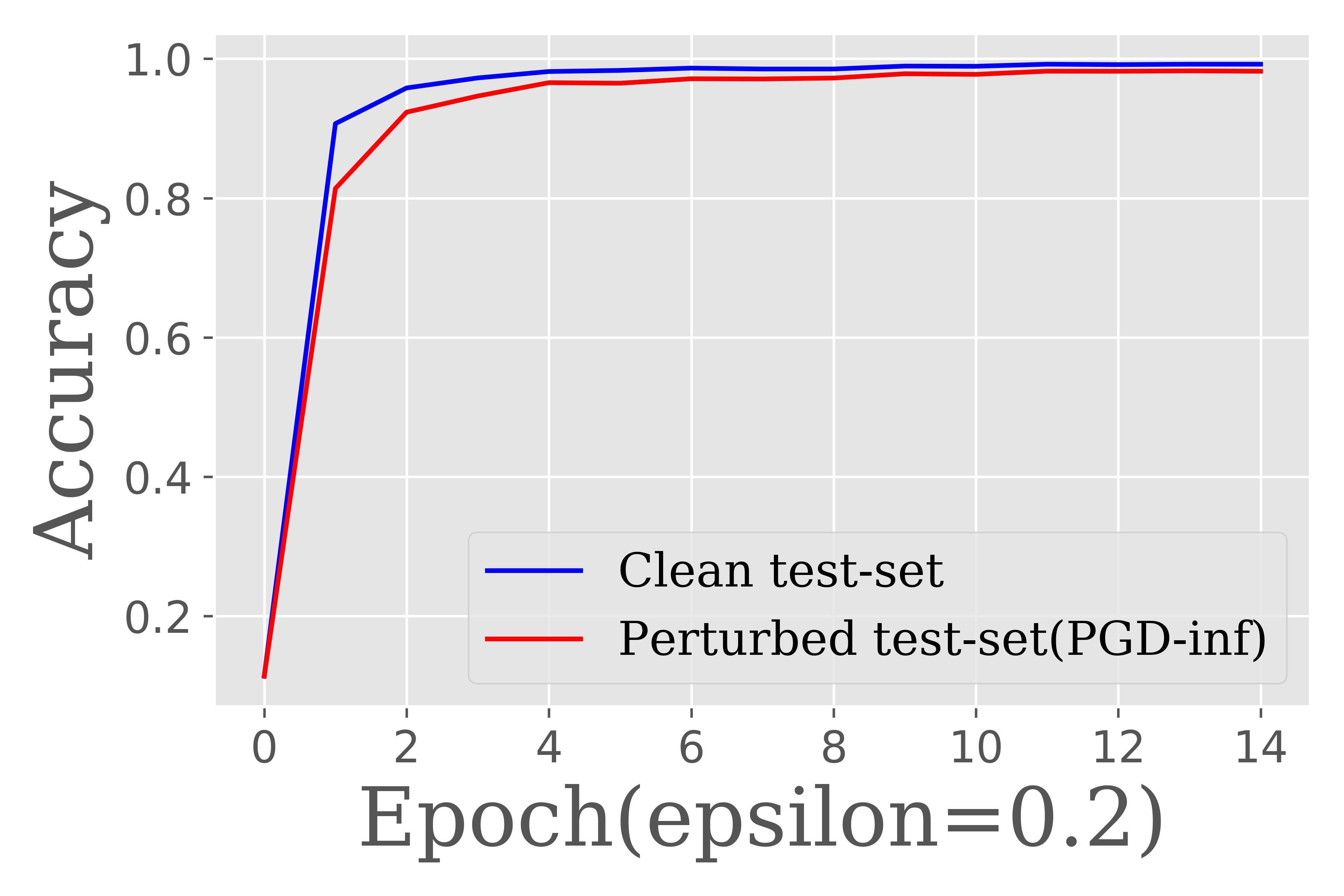}
    \endminipage\hfill
    \caption{Accuracy Curves on clean MNIST test-set and perturbed test-set during \emph{adv.FGSMR} training process. Perturbed test-set are generated by \emph{PGD-inf} attack with $\epsilon=0.1$. Upper figure shows the accuracy curve for \emph{adv.FGSMR} with $\epsilon=0.1$; Lower figure shows the accuracy curve for \emph{adv.FGSMR} with $\epsilon=0.2$.}
    \label{fig:M_curves}
\end{figure}

\paragraph{Model Architecture}
Model architecture for MNIST is composed of four convolution layers and two fully-connected layers. Detailed parameters are showed in Table~\ref{tb:model structure}.
\begin{table}[htb]
\centering  
\begin{tabular}{ccccc} \\
\thickhline
Layer  &Kernel & Step & Padding & output\\
       &size &size & Padding & dimension\\
\hline
Conv1  &3$\times$3 & 1 & 1 & 32\\
Conv2  &3$\times$3 & 2 & 1 & 32\\
Conv3  &3$\times$3 & 1 & 1 & 64\\
Conv4  &3$\times$3 & 2 & 1 & 64\\
fc5    &- &-&-&100 \\
fc6    &- &-&-& 10 \\
\thickhline
\end{tabular}
\caption{Model architecture on MNIST.}
\label{tb:model structure}
\end{table}
\paragraph{Training Curves}
The training is based on \emph{SGD} optimizer. Whole training process contains 15 epochs. Learning rate is set to 0.1 for the first 10 epochs and 0.01 for the last 5 epochs. $\lambda$ is set to 0.2 for the curvature regularization. Figure~\ref{fig:M_curves} shows accuracy curves for \emph{adv.FGSMR} with $\epsilon=0.1, 0.2$ respectively. 
\subsection{CIFAR-10}
\begin{figure*}[htb]
    \centering
     \minipage{0.33\textwidth}
        \includegraphics[width=\linewidth]{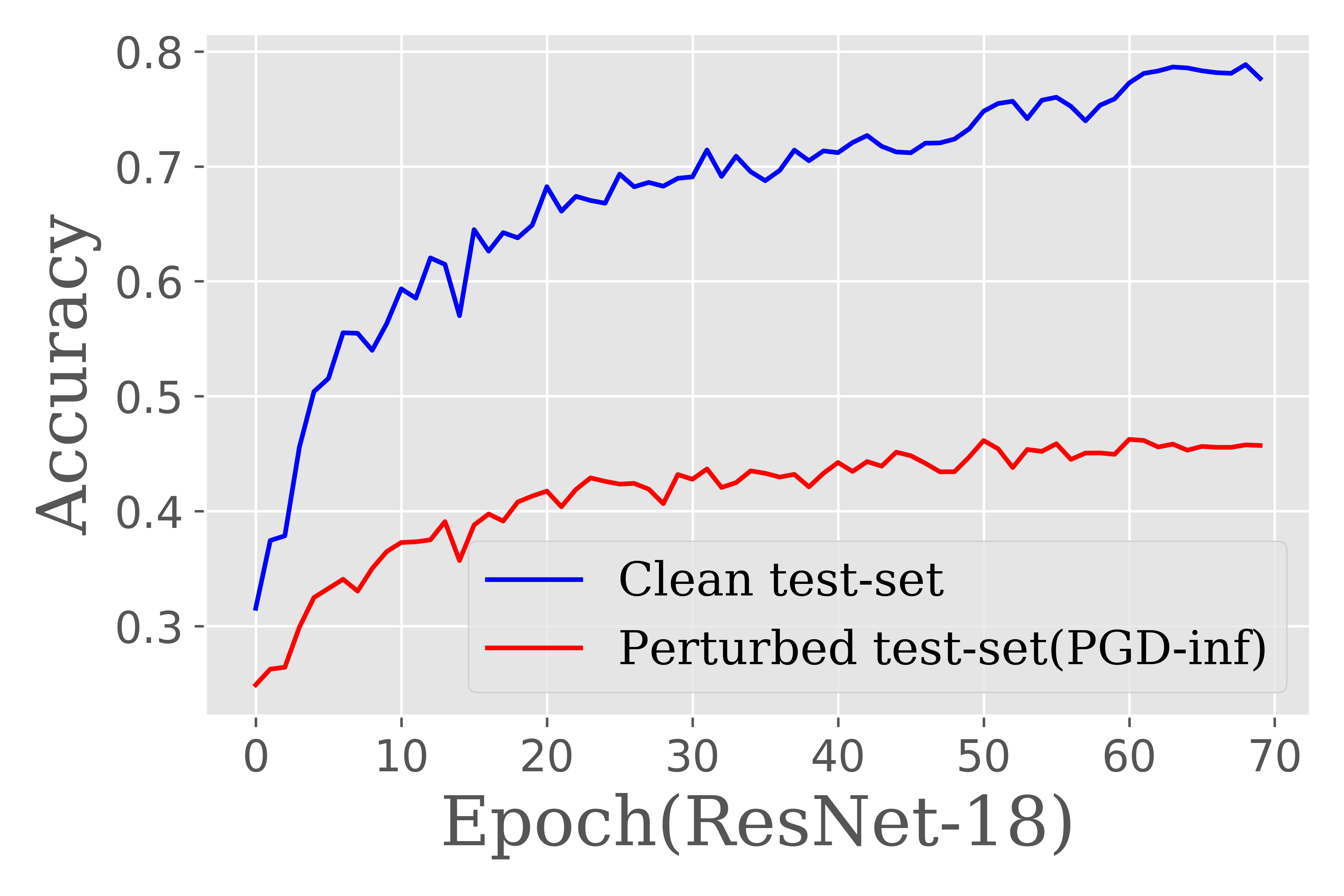}
    \endminipage\hfill
    \minipage{0.33\textwidth}
        \includegraphics[width=\linewidth]{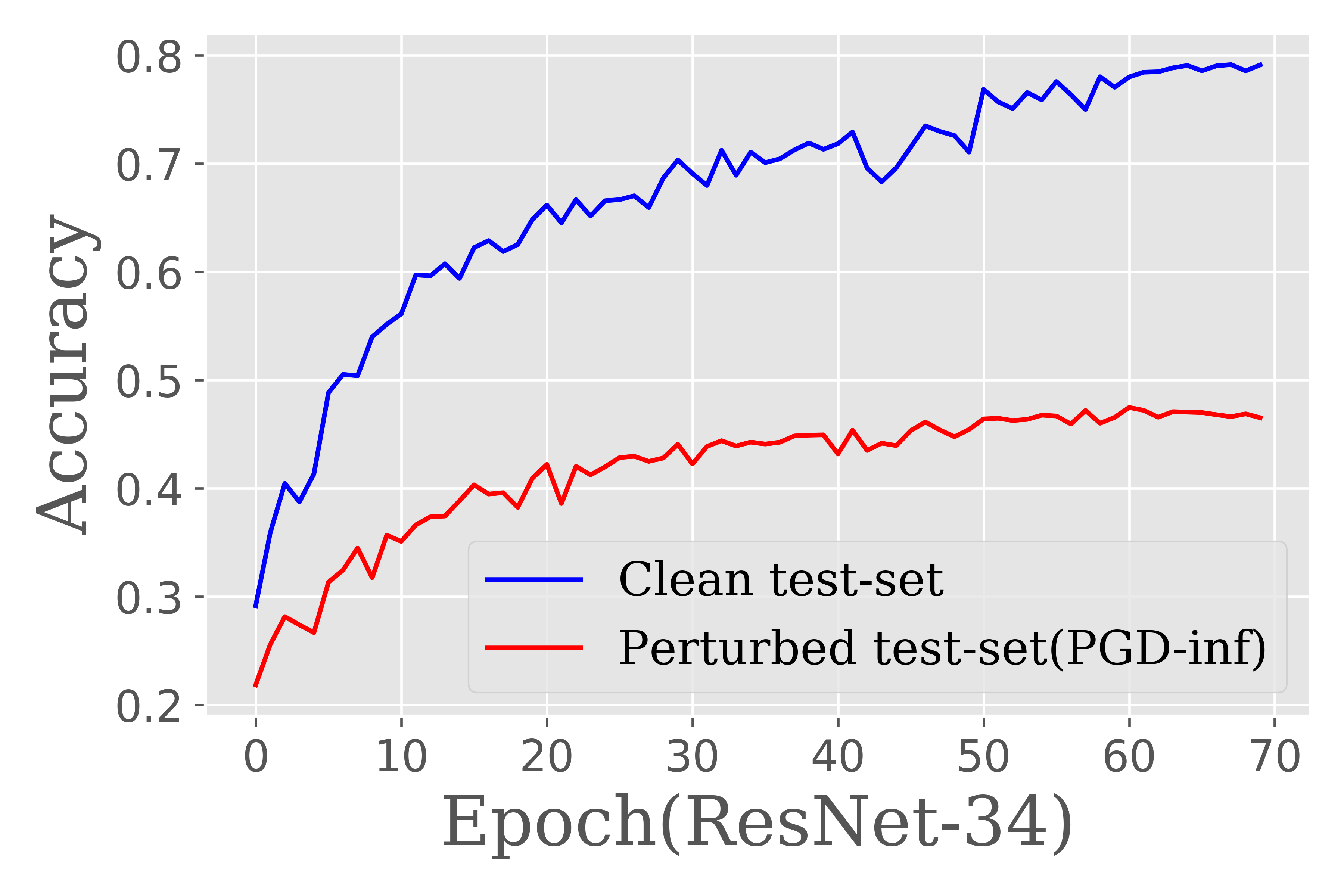}
    \endminipage\hfill
    \minipage{0.33\textwidth}
        \includegraphics[width=\linewidth]{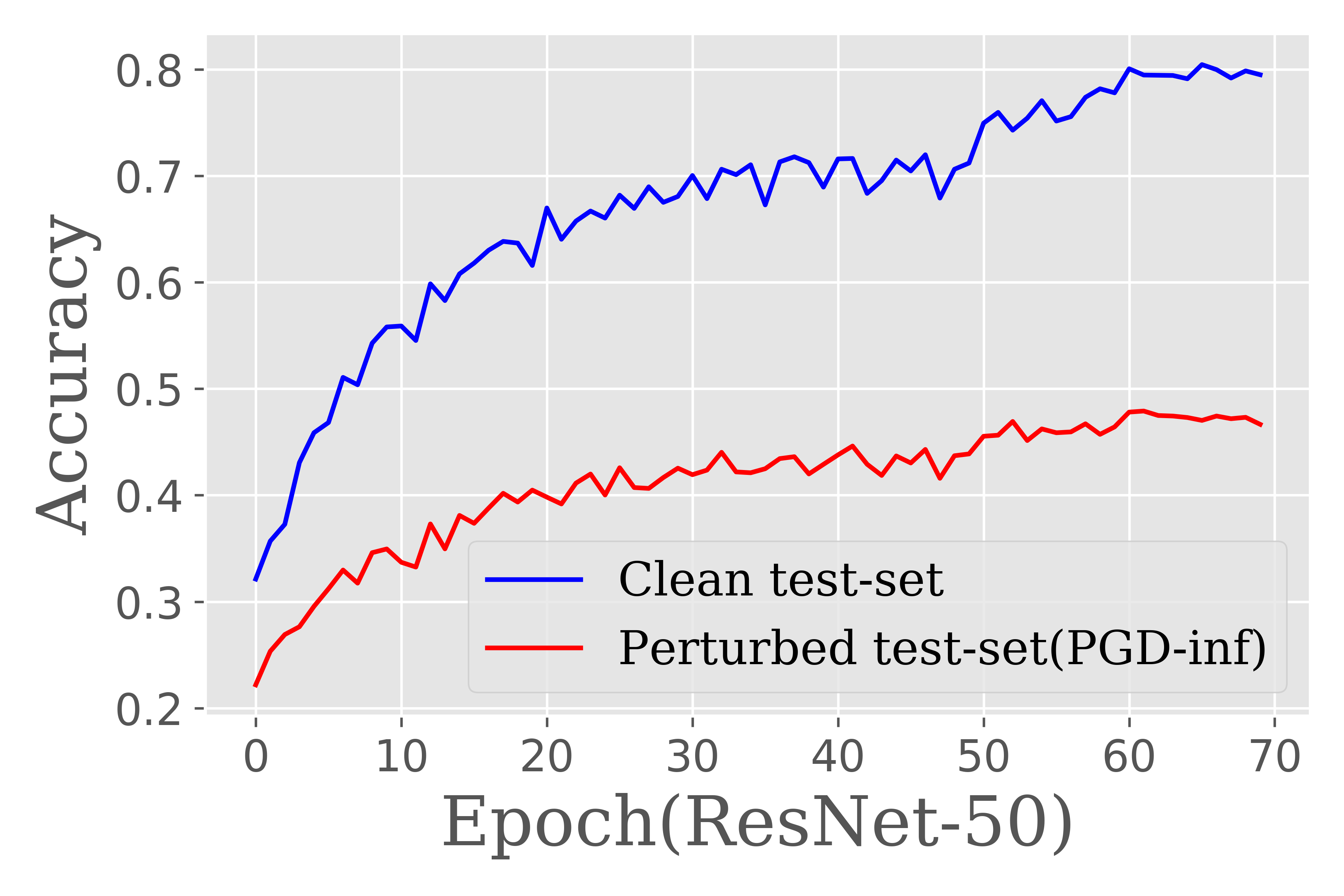}
    \endminipage\hfill    
    \\
    \minipage{0.33\textwidth}
        \includegraphics[width=\linewidth]{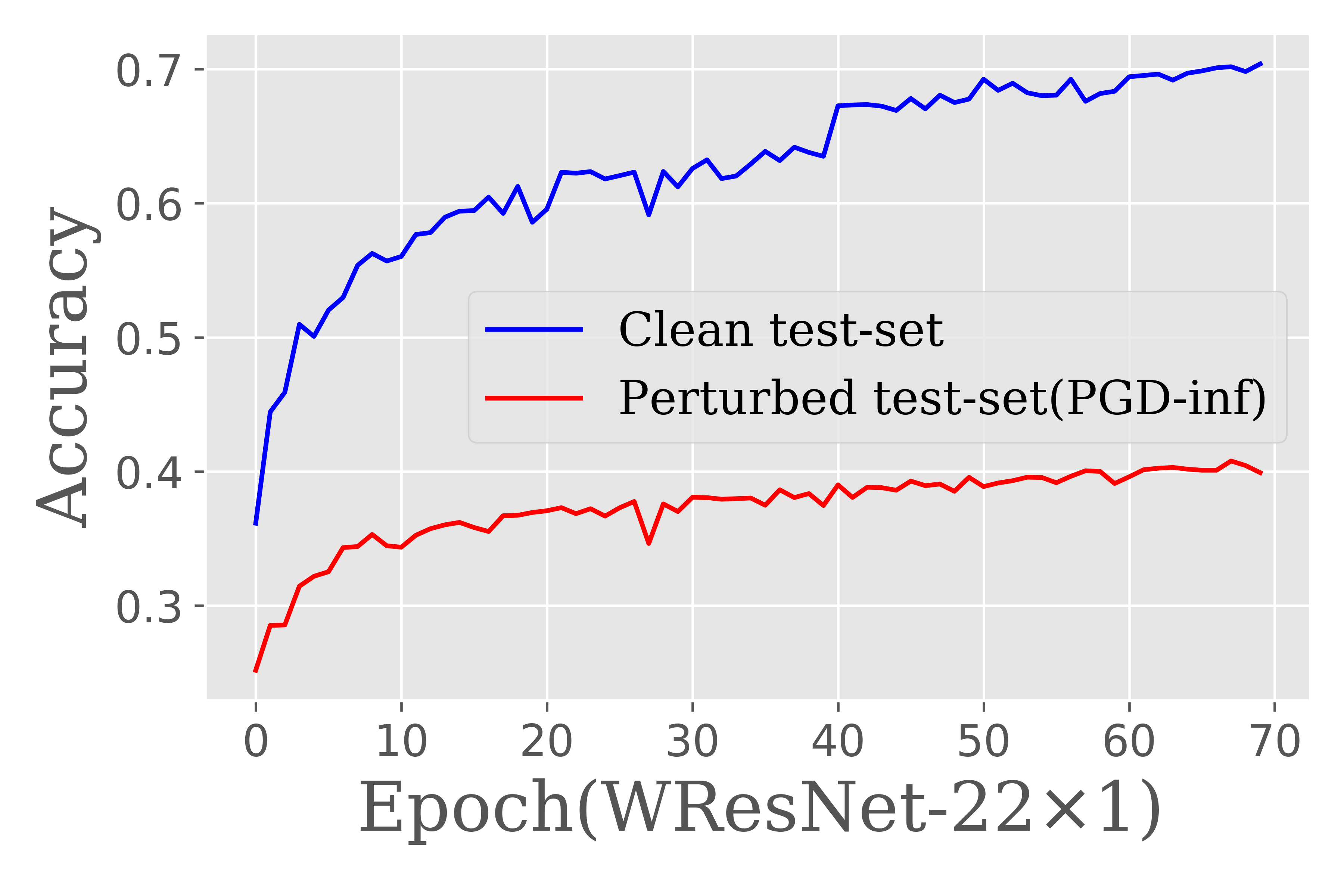}
    \endminipage\hfill
    \minipage{0.33\textwidth}
        \includegraphics[width=\linewidth]{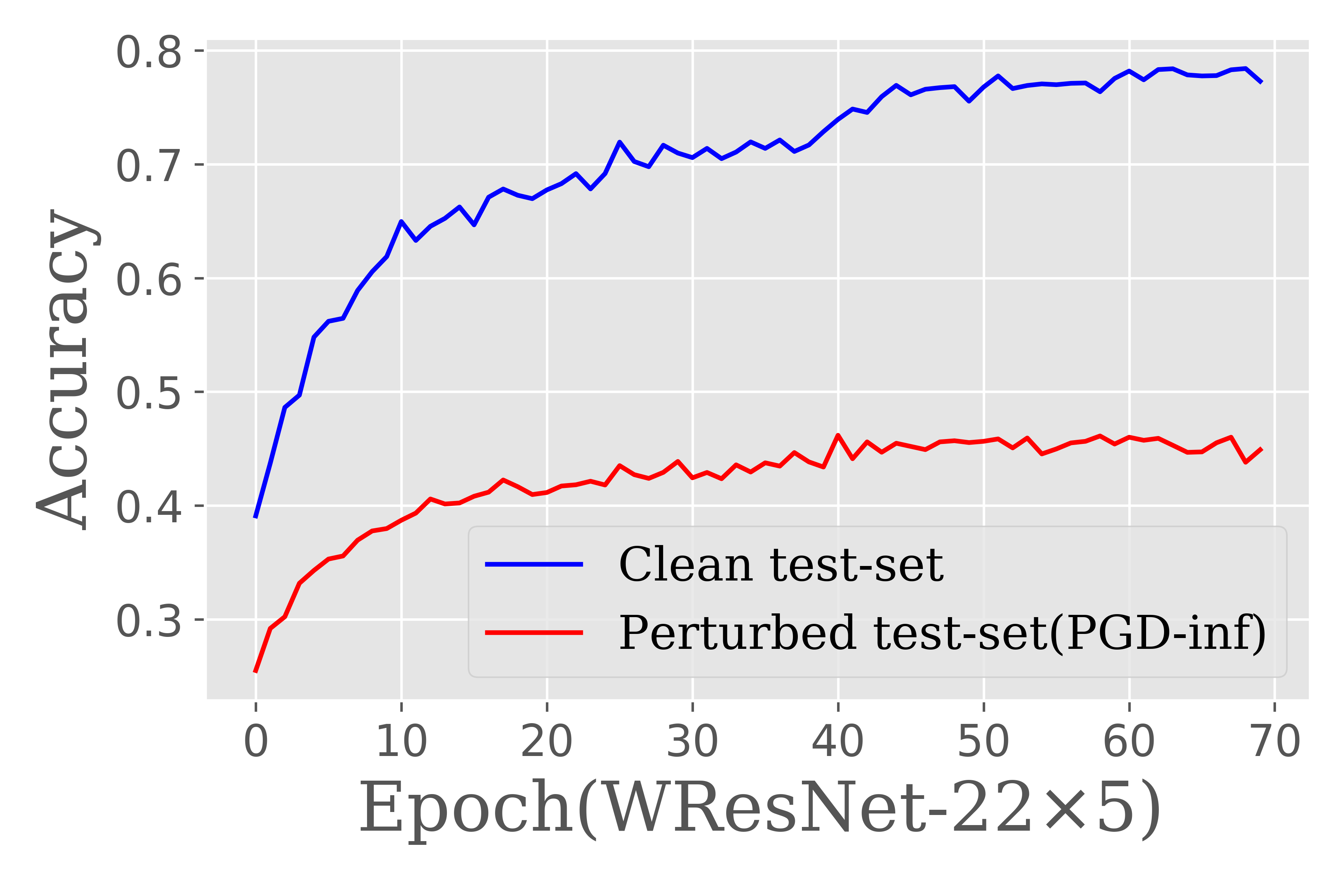}
    \endminipage\hfill
    \minipage{0.33\textwidth}
        \includegraphics[width=\linewidth]{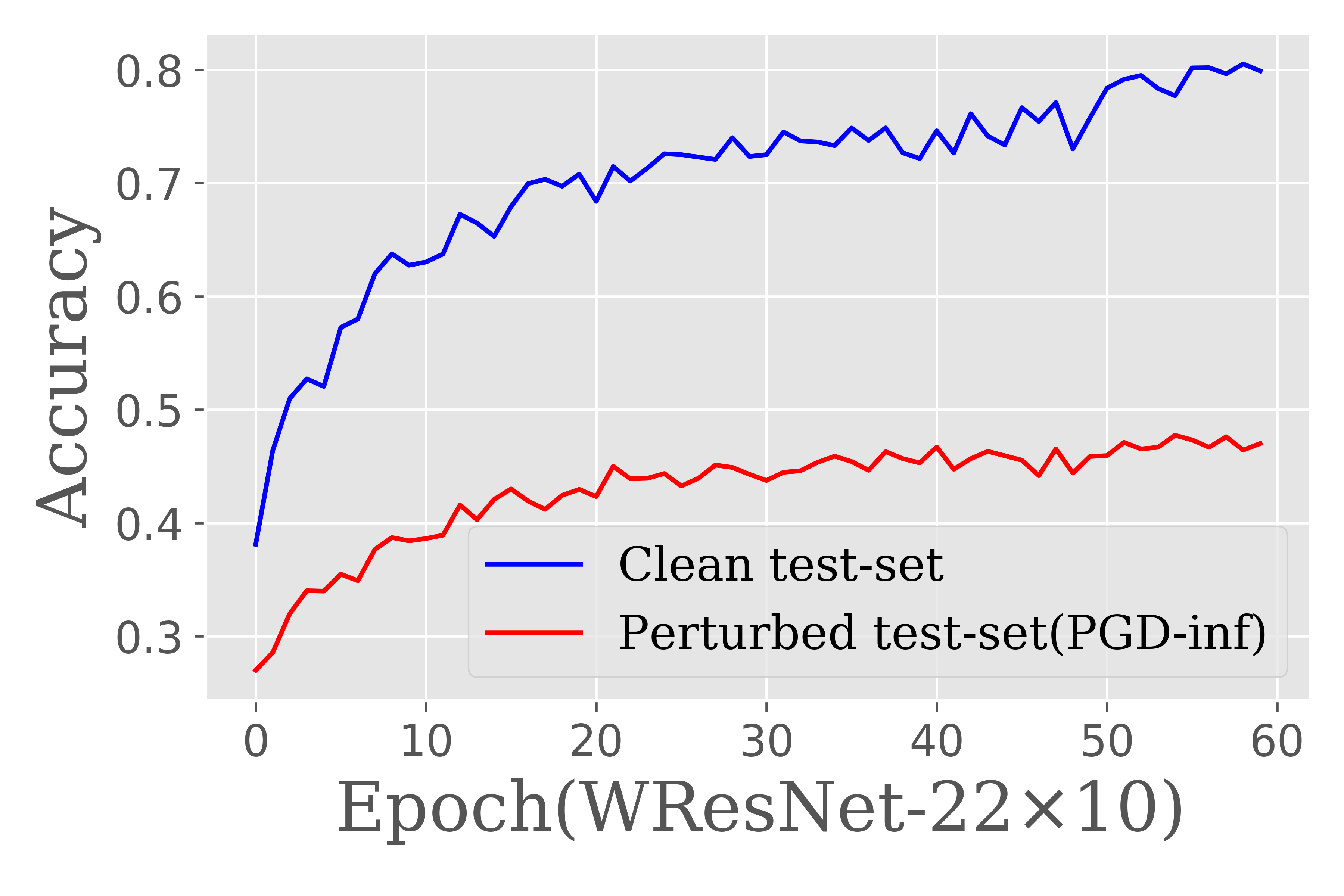}
    \endminipage\hfill
    \caption{The Accuracy curves on clean and perturbed test-set for various robust models with \emph{adv.FGSMR} ($\epsilon=8.0/255$). The perturbed test-set is generated by \emph{PGD-inf} attack with $\epsilon=8.0/255$. }
    \label{Fig:CIFAR_CURVES}
\end{figure*}

\paragraph{Training Curves}
\emph{Adam} optimizer is used for training models on CIFAR-10. Besides, learning rate decay is adopted during training process. Detailed training settings please refer to Table~\ref{tb:Training_strategy}. Figure~\ref{Fig:CIFAR_CURVES} shows the accuracy curves for training \emph{ResNet-18}, \emph{ResNet-34}, \emph{ResNet-50}, \emph{Wide ResNet-22$\times$1$\times$0$\times10$}, \emph{Wide ResNet-22$\times$5$\times$0$\times10$} and \emph{Wide ResNet-22$\times$10$\times$0$\times10$} respectively with \emph{adv.FGSMR} ( $\epsilon=8.0/255$).

\begin{table}[htb]
\centering  
\begin{tabular}{ccccc} \\
\thickhline
Models  &Learning rate & $\lambda$\\
\hline
\multirow{3}{*}{ResNet-18}  
&0-50:$1e^{-3}$ & 0.5\\
&50-60:$5e^{-4}$ & 0.3\\
&60-70:$1e^{-4}$ & 0.3\\
\hline
\multirow{3}{*}{ResNet-34}  
&0-50:$1e^{-3}$ & 0.5  \\
&50-60:$5e^{-4}$ & 0.4\\
&60-70:$1e^{-4}$ & 0.4\\
\hline
\multirow{3}{*}{ResNet-50}  
&0-50:$1e^{-3}$ & 0.5  \\
&50-60:$5e^{-4}$ & 0.4\\
&60-70:$1e^{-4}$ & 0.4\\
\hline
\multirow{3}{*}{WResNet-22$\times$1}  
&0-40:$1e^{-3}$ & 0.5  \\
&40-50:$5e^{-4}$ & 0.3\\
&50-70:$1e^{-4}$ & 0.3\\
\hline
\multirow{3}{*}{WResNet-22$\times$5}    
&0-40:$1e^{-3}$ & 0.5  \\
&40-50:$5e^{-4}$ & 0.4\\
&50-70:$1e^{-4}$ & 0.4\\
\hline
\multirow{3}{*}{WResNet-22$\times$10}    
&0-40:$1e^{-3}$ & 0.5  \\
&40-50:$5e^{-4}$ & 0.4\\
&50-60:$1e^{-4}$ & 0.4\\
\thickhline
\end{tabular}
\caption{Detailed settings for training robust models with \emph{adv.FGSMR}. $\lambda$ is the parameter for curvature regularization.}
\label{tb:Training_strategy}
\end{table}